\documentclass[usenatbib,letter]{mn2e} 
\usepackage{amsmath} 
\usepackage{amssymb} 
\usepackage{graphics}
\usepackage{epsfig}  
\def\be{\begin{equation}}
\def\ee{\end{equation}}
\def\ba{\begin{eqnarray}}
\def\ea{\end{eqnarray}}

\newcommand{\beq}{\begin{eqnarray}}  
\newcommand{\eeq}{\end{eqnarray}}  
  
\newcommand{\apj}{ApJ}  
\newcommand{\apjs}{ApJS}  
\newcommand{\apjl}{ApJL}  
\newcommand{\aj}{AJ}  
\newcommand{\mnras}{MNRAS}  
\newcommand{\aap}{A\&A}

\newcommand{\pasp}{PASP}    
\newcommand{\pasj}{PASJ}    
\newcommand{\avg}[1]{\langle{#1}\rangle}  
\newcommand{\hMpc}{{\ifmmode{h^{-1}{\rm Mpc}}\else{$h^{-1}$Mpc }\fi}}  
\newcommand{\hGpc}{{\ifmmode{h^{-1}{\rm Gpc}}\else{$h^{-1}$Gpc }\fi}}  
\newcommand{\hmpc}{{\ifmmode{h^{-1}{\rm Mpc}}\else{$h^{-1}$Mpc }\fi}}  
\newcommand{\hkpc}{{\ifmmode{h^{-1}{\rm kpc}}\else{$h^{-1}$kpc }\fi}}  
\newcommand{\hMsun}{{\ifmmode{h^{-1}{\rm {M_{\odot}}}}\else{$h^{-1}{\rm{M_{\odot}}}$}\fi}}  
\newcommand{\hmsun}{{\ifmmode{h^{-1}{\rm {M_{\odot}}}}\else{$h^{-1}{\rm{M_{\odot}}}$}\fi}}  
\newcommand{\Msun}{{\ifmmode{{\rm {M_{\odot}}}}\else{${\rm{M_{\odot}}}$}\fi}}  
\newcommand{\msun}{{\ifmmode{{\rm {M_{\odot}}}}\else{${\rm{M_{\odot}}}$}\fi}}  

\begin{document}

\title[High-z UV: hints for a clumpy ISM?]{Simulated vs. observed UV emission at high redshift: a hint for a clumpy ISM?}
\author[Forero-Romero et al.]{
\parbox[t]{\textwidth}{\raggedright 
  Jaime E. Forero-Romero$^1$ \thanks{Email: jforero@aip.de},
  Gustavo Yepes$^2$, 
  Stefan Gottl\"ober$^1$,
  Steffen R. Knollmann$^2$, 
  Arman Khalatyan$^3$, 
  Antonio J. Cuesta$^4$, 
  Francisco Prada$^4$}\\
\vspace*{6pt}\\
$^1$Astrophysikalisches Institut Potsdam, An der Sternwarte 16, 14482 Potsdam, Germany\\ 
$^2$Grupo de Astrof\'{\i}sica, Universidad Aut\'onoma de Madrid,   Madrid
E-28049, Spain\\
$^3$Laboratoire d'Astrophysique de Marseille, Observatoire Astronomique de Marseille Provence, \\
$\ $Technopole de l'\'Etoile - Site de Chateau-Gombert, 38 rue Fr\'ed\'eric Joliot-Curie, 13388 Marseille C\'edex 13, France\\
$^4$Instituto de Astrof\'{\i}sica de Andaluc\'{\i}a (CSIC), Camino Bajo de Hu\'etor 50, E-18008, Granada, Spain
}
\maketitle

\begin{abstract}

We discuss the rest-frame UV emission between $5< z < 7$ from
 the {\em MareNostrum High-z Universe}, a SPH simulation done 
 with more than 2 billion particles. Cosmological simulations of galaxy formation generally overpredict  the UV restframe
  luminosity function at high redshift, both at the bright and faint ends.
In this Letter we explore a
  dust attenuation model where a larger extinction is applied to 
  star  populations younger than a given age, mimicking the effect
  of a clumpy interstellar medium. We show that this  scenario fits
  reasonably well 
both the   UV luminosity functions and the UV-continuum  slopes derived  from
  observations.
The model assumes  a large obscuration for  stars younger than $25$ Myr 
from the gas clouds where they should be embedded at their formation
time. We find that the optical depth in these clouds should be between
$30$ and $100$ times larger than the mean optical depth for the
homogeneous part of the interstellar medium. These values  are one
order of magnitude larger than those estimated in local galaxies. 
Therefore,  we conclude that $\Lambda$CDM  predictions for the
  high-z UV emission   can  accommodate 
 the current observations if we consider 
 a  dust extinction model  based on the assumption of a clumpy 
  environment at high redshift.

\end{abstract}


\section{Introduction}

One of the most relevant issues in cosmology is understanding the evolution of
galaxy populations. Thanks to recently developed techniques and improved instrumentation, observations of
high redshift galaxies ($z>5$) began to be possible during the
last years
\citep{iwata07,2007ApJ...670..928B,bouwens08,2009MNRAS.395.2196M,2009ApJ...697.1493S,2009arXiv0908.3191O,2009arXiv0909.1803B,2009arXiv0910.0001B,2009arXiv0909.1806O,2009arXiv0909.2255B,2009arXiv0909.2437M}.

The fundamental quantity used in the exploration of these galaxy populations is the
galaxy luminosity function (LF). In particular, the LF in the
rest-frame UV provides valuable information on key physical processes. The UV
output of a galaxy imposes constraints on the instantaneous formation rate and
amount of young stars in a galaxy. This kind of census provides information on
the role of galactic star populations in the reionization of the Universe
\citep{2003MNRAS.342..439S}.

Observationally, the method of choice to gather a large sample of galaxies at
high redshift is the drop-out technique. This technique detects the spectral
discontinuity at $912$ \AA\  (Lyman break) due to the absorption of UV
radiation by
intergalactic hydrogen gas using multi-wavelength broad band imaging
\citep{1996AJ....112..352S}.
 The galaxies discovered using this method receive the name of 
Lyman Break Galaxies (LBGs).

Motivated by the observational results, recent theoretical studies of LBGs beyond
$z>5$ have been developed \citep{2006MNRAS.366..705N}. These theoretical studies,
usually require computational techniques to produce a mock galaxy population which can
be directly compared with the observed one. In the case of
the rest-frame UV LF, two important elements are required in the model:
the stellar population (masses, ages and metallicities) and the
corresponding gas content in order to derive the extinction properties.
Modelling the intrinsic UV emission through stellar population synthesis
models is now a standard procedure \citep{1998MNRAS.300..872M}. On the
other hand, having a proper estimation of the extinction is a more
uncertain process.

The estimation of the extinction can assume the form of a simple
Calzetti law \citep{2000ApJ...533..682C}, where usually the UV magnitudes of
each galaxy are reduced by the same amount regardless of the physical
properties of the gas in the galaxy \citep{2006MNRAS.366..705N}. If the galaxy model
provides information on the gas and its metallicity contents, it would
be desirable to implement a more physical model for the dust
obscuration and try to put some constraints on the amount and
distribution (i.e. clumpiness) of the dust on galactic scales
\citep{2005MNRAS.359..171I}.

In this Letter we present the first results from the  {\em MareNostrum High-z Universe}. The simulation follows self-consistently
the dynamical evolution of more than 2 billion dark matter and gas
particles and includes a prescription for star formation, supernovae
feedback and UV background.  Here, we will present the numerical
estimates for the continuum rest-frame UV emission from the stellar
component, and a dust attenuation model. Our main objective is
to compare the numerical predictions with the observed evolution of the
UV LF between redshifts $z=5$ and $z=7$.

\section{Simulation and Galaxy Finding}
\label{sec:simulation}
The {\em MareNostrum  High-z Universe} simulation{\footnote{\tt http://astro.ft.uam.es/marenostrum}} follows the non linear evolution of structures in  
baryons (gas and stars) and dark matter, starting  from $z= 60$
 within a cube of $50\hMpc$ comoving on a side. 
The cosmological parameters used correspond to WMAP1 data \citep{2003ApJS..148..175S} and are $\Omega_{\rm m}=0.3$, $\Omega_{\rm b}=0.045$, $\Omega_\Lambda=0.7$, $\sigma_8=0.9$, a Hubble parameter $h=0.7$, and a spectral index $n=1$. The initial density field has been sampled by $1024^3$ dark
matter particles with a mass of $m_{\rm DM} = 8.2 \times 10^6 \hMsun $ and
$1024^3$ SPH gas particles with a mass of $m_{\rm gas} = 1.4 \times 10^6 \hMsun$. The
simulation has been performed using the TREEPM+SPH code \texttt{GADGET-2}
\citep{Springel05}.  Radiative and Compton cooling
processes for an optically thin primordial plasma  of Helium and Hydrogen
are included. We assumed photo-ionisation by an external spatially
uniform UV-background adopting \citet{Haardt96}.  
The  physics of star formation is treated by means of a sub-resolution
model in which the gas of the interstellar medium (ISM) is described as a
multiphase medium of hot and cold gas \citep{yepes:97,springel:03}. Stars can
be formed in regions that are sufficiently dense and cold. 
We consider the thermal feedback of supernovae as well as the effects
of stellar winds following the model described in  \citet{springel:03}
The gravitational smoothing scale  was set to 2 \hkpc in comoving
coordinates.  The simulation was run in the MareNostrum supercomputer
using  800 processors  simultaneously. This simulation is intended  to study the
early phases of galaxy formation. It is currently at z=4.55, 
after spending more than 4.5 million cpu hours.

We identify the objects  in the simulations 
using the \texttt{AMIGA} Halo Finder (\texttt{AHF})\footnote{It is
  a MPI+OpenMP hybrid halo finder to be downloaded freely from
{\tt http://www.popia.ft.uam.es/AMIGA}}  which identifies both halos and
  subhalos. \texttt{AHF} is an improvement of the \texttt{MHF} halo finder
\citep{Gill2004}, which locates local overdensities in an adaptively smoothed
density field as prospective halo centers. \texttt{AHF} is described in detail
in \citet{Knollmann2009}.

In \texttt{AHF} the local potential minima are computed for each of these
density peaks and the gravitationally bound particles are determined.
Only peaks with at least 20 bound particles are considered as haloes and
retained for further analysis in \texttt{AHF}. The thermal energy of gas particles is taken into account during the calculation of the binding energy.

All objects with more than  $1000$ particles,  dark matter, gas and
stars combined, are used in our present analyses. We assume a  galaxy  is
resolved  if the object  contains   $200$ or more star particles, 
which  corresponds  to objects with $\gtrsim 400$
particles of gas. This   ensures  a proper estimation of the
average gas column densities in our numerical galaxies.
We  then build the  full Spectral Energy Distribution (SED)  and
compute the dust attenuation using the information from the 
stars and gas particles of each galaxy. In  the following 
section we  explain  the details of these procedures.

\section{Spectral Modelling}
\label{sec:model}

The photometric properties of the galaxies are calculated employing the
stellar population synthesis model \texttt{STARDUST}
\citep{1999A&A...350..381D}, using the methods described in
\cite{2003MNRAS.343...75H}. We adopt a Salpeter Initial Mass Function
(IMF). The SEDs are built using the \texttt{AHF} catalogs. 
Nevertheless, we have checked that all the results quoted in this Letter remain valid 
if we identify galaxies   by the  Friend-of-Friends algorithm 
that run over the star particles only. 

The dust attenuation model parametrizes both the extinction in a homogeneous
interstellar medium (ISM) and the molecular clouds around young stars,
following the physical model of \cite{2000ApJ...539..718C}. The attenuation
from dust in the homogeneous ISM assumes later a slab geometry, while the
additional attenuation for young stars assumes a spherical symmetry.

We describe first the optical depth for the homogeneous interstellar
medium, denoted by $\tau_{\lambda}^{H}$. We take the mean
perpendicular optical depth of a galactic disc at wavelength $\lambda$
to be 

\begin{equation}
\tau_{\lambda}^{H}  = \eta
\left(\frac{A_{\lambda}}{A_{V}}\right)_{Z_{\odot}}\left(\frac{Z_g}{Z_{\odot}}\right)^r\left(\frac{\avg{N_{H}}}{2.1
  \times 10^{21} \mathrm{atoms\ cm}^{-2}}\right),
\label{eq:ISM}
\end{equation}
where $A_\lambda/A_V$ is the extinction curve from \cite{1983A&A...128..212M}, $Z_{g}$ is the gas metallicity, $\avg{N_H}$ is the mean atomic hydrogen
column density and $\eta=(1+z)^{-\alpha}$ is a factor that takes into account the evolution of the dust to gas ratio at different redshifts,
 with $\alpha>0$ from the available constraints based on simplified theoretical models \citep{2003PASJ...55..901I} and observations around $z\sim 3$ \citep{2006ApJ...644..792R}. Given the uncertainties in this value, we will treat $\alpha$ as a free parameter.

The mean H column density is calculated as

\begin{equation}
\avg{N_H} = X_H \frac{M_{g}}{m_{p}\pi r_{g}^2}\mathrm{atoms\ cm}^{-2},
\end{equation}

where $X_H = 0.75$ is the universal mass fraction of Hydrogen, $M_g$ is the
mass in gas, $r_{g}$ is the radius of the galaxy and $m_p$ is the proton
mass. The extinction curve depends on the gas
metallicity $Z_{g}$ and is based on an interpolation between the solar
neighborhood and the Large and Small Magellanic Clouds ($r=1.35$ for $\lambda
< 2000 $\AA\ and $r=1.6$ for $\lambda > 2000$\AA).

The radius,  stellar and gas masses for each galaxy  are taken from
the \texttt{AHF} catalogs.  Computing $\avg{N_H}$ from the galaxy
catalogs turns out to give  similar results,  on average,  than
integrating the 3D gas distribution of the galaxies using the
appropriate SPH kernel.  The difference between the average Hydrogen
column densities  calculated in these two different ways are less than
$15\%$ regardless  of the size of the  galaxy,  provided that the gas
distribution  is sampled with more than $200$  particles.

In addition to the foreground, homogeneous ISM extinction, we also
model, in a simple manner,  the attenuation of young stars that are 
 embedded  in their nest  clouds. Stars
younger than a given age, $t_{c}$, are subject to an additional attenuation with
mean perpendicular optical depth

\begin{equation}
\tau_{\lambda}^{C} = \left(\frac{1}{\mu}-1\right) \tau_{\lambda}^{H},
\label{eq:young}
\end{equation}
where $\mu$ is the fraction of the total optical depth seen by the young
stars that is found in the homogeneous ISM. \cite{2004MNRAS.349..769K} in a
study of 115 nearby galaxies, spanning a wide range of star formation
activities, find that $\mu$ is typically $\sim 1/3$ with a wide range of
scatter between $0.1$ and $0.6$. Given the possible uncertainties in extrapolating the typical values
of $\mu$ to high redshift, we will treat this quantity as a free
parameter.

\section{Analysis and results}

\label{sec:analysis}

\begin{figure*}
\begin{center}
\includegraphics[width=0.33\textwidth]{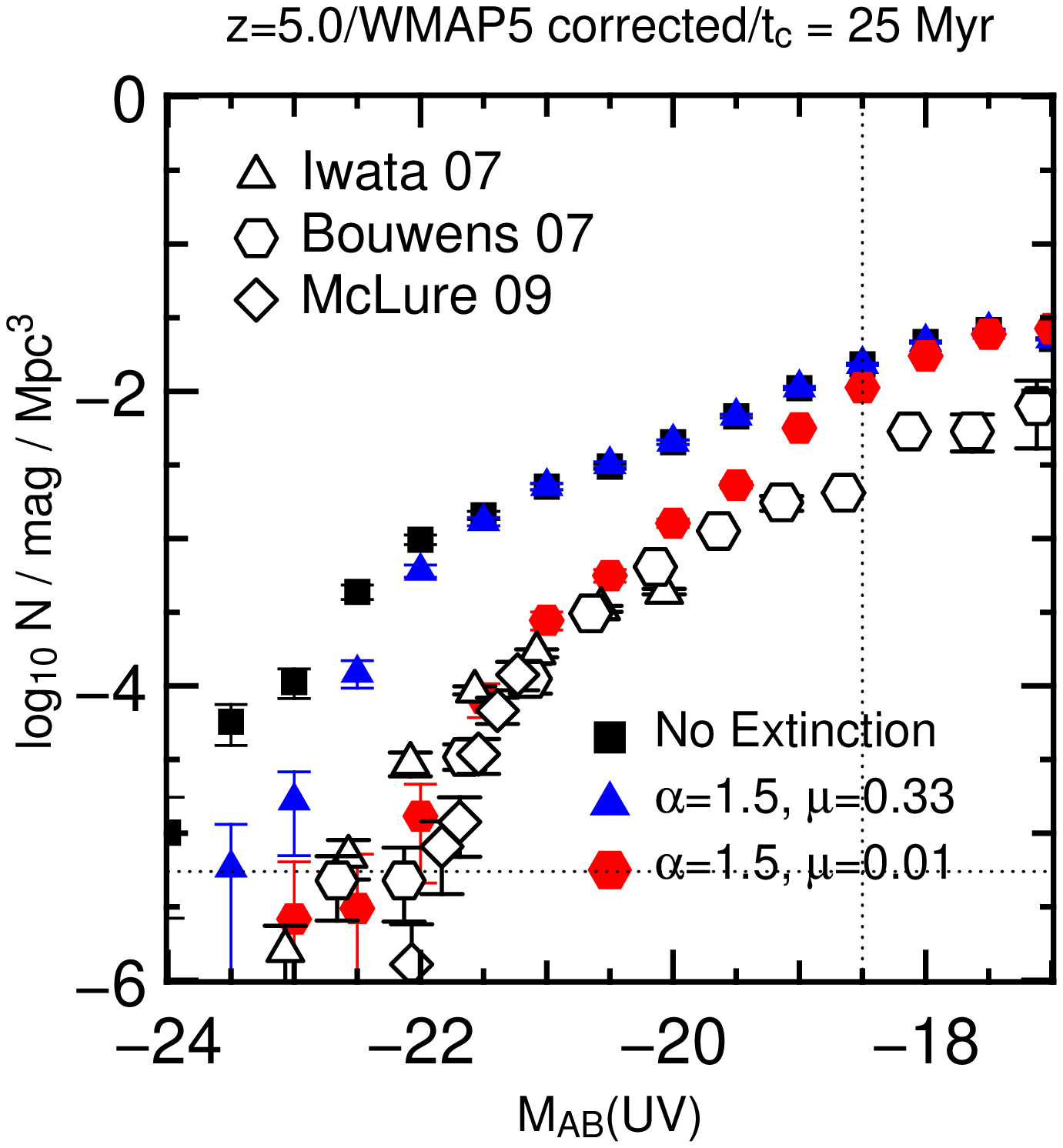}\hspace{-0.9cm}
\includegraphics[width=0.33\textwidth]{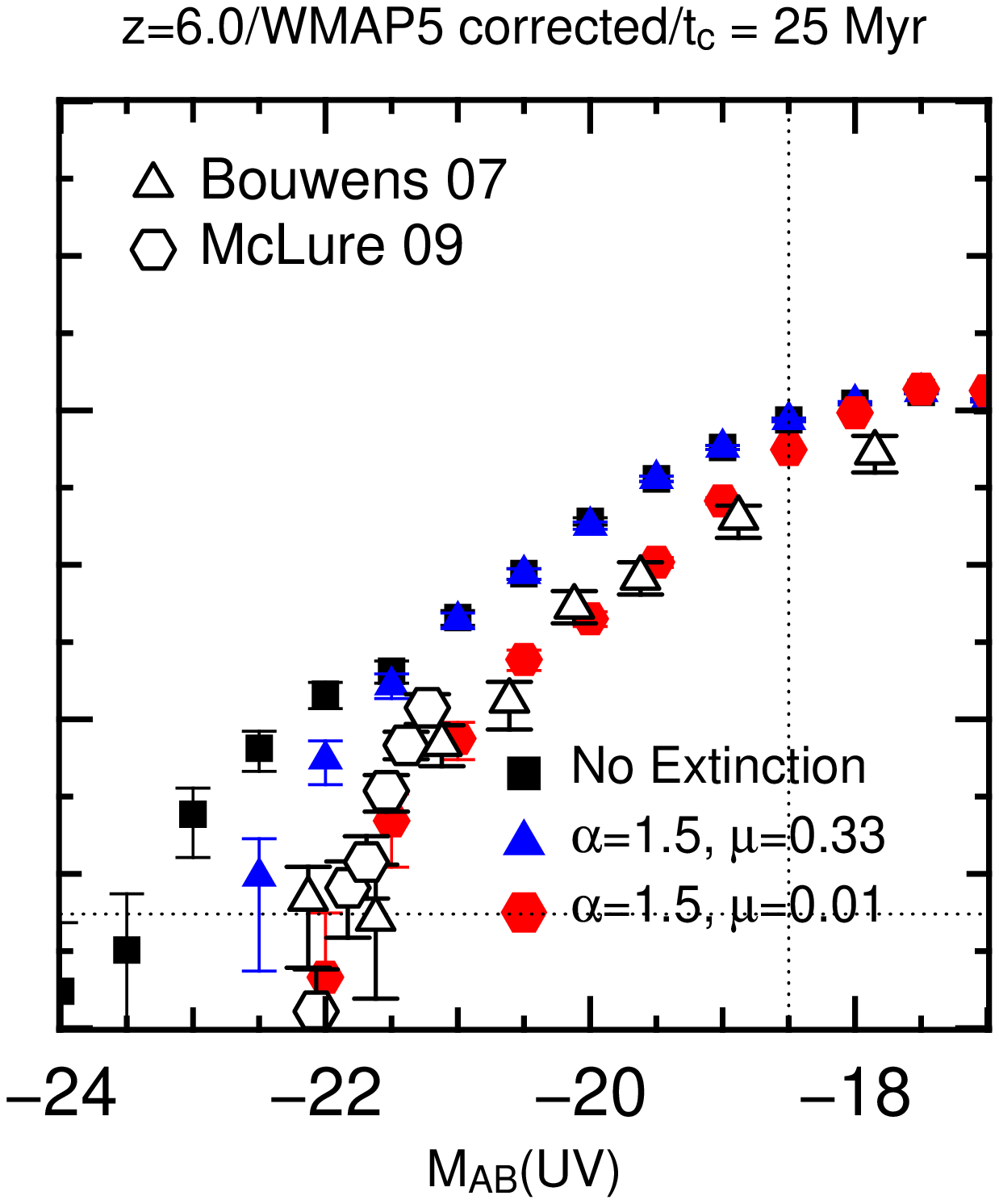}\hspace{-0.9cm}
\includegraphics[width=0.33\textwidth]{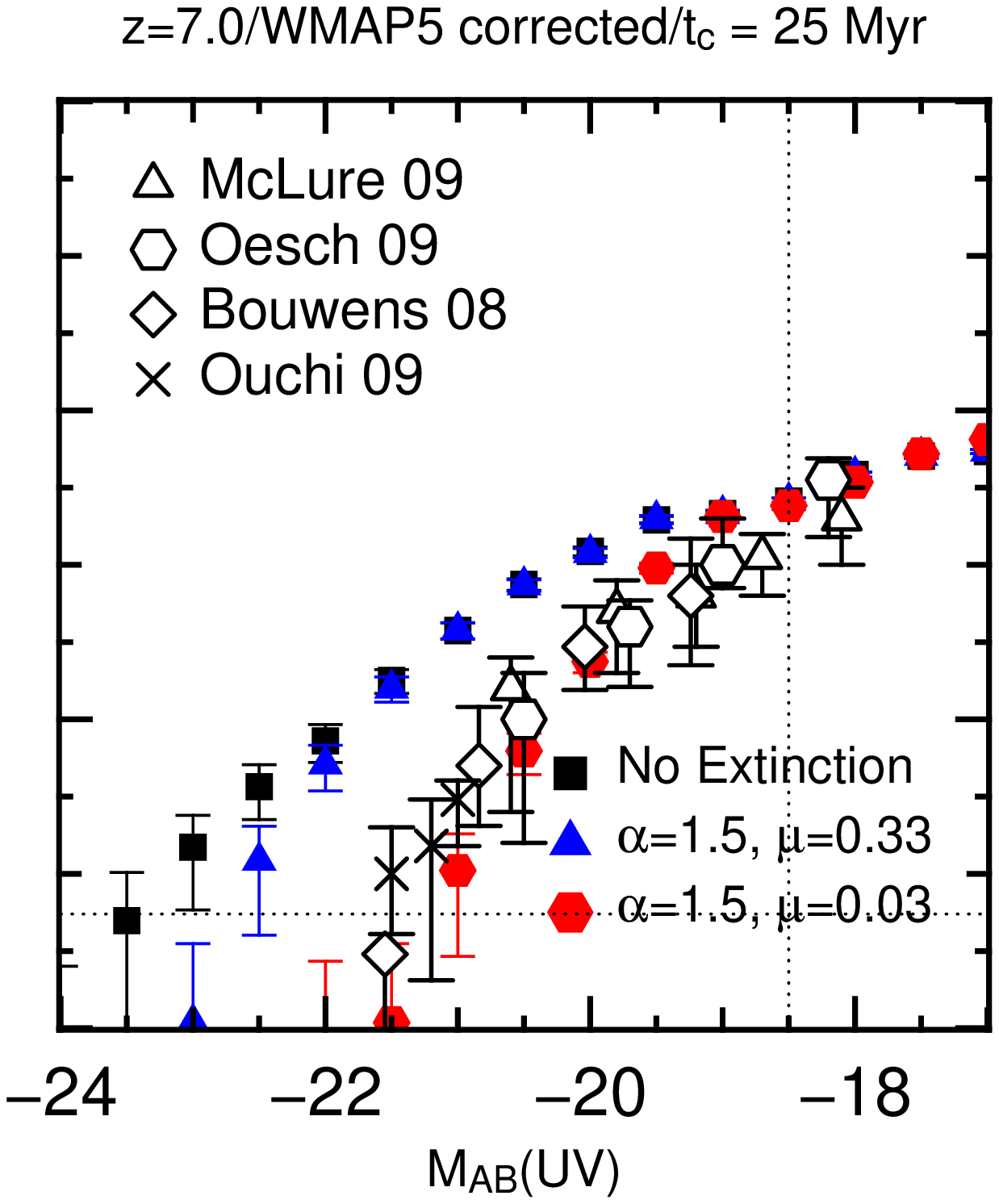}
\end{center}
\caption{Luminosity functions at redshifts $z\sim$ $5$, $6$ and $7$. The error bars are poissonian. The horizontal line shows the limits of one
  object per unit magnitude in the simulation volume, for the used bin size 
  $\Delta M_{\rm AB}=0.50$. The vertical line
  indicates the approximate limit for well resolved objects following a
  criteria of more than $200$
  star particles. This limit corresponds to similar UV magnitudes
  at every redshift. 
  The simulation data has been   already corrected 
  for the different abundance of dark matter halos between the WMAP1 and WMAP5
  cosmologies. The open symbols represent the observations. The filled symbols
  represent different   extinction prescriptions on the simulated galaxies: no
  extinction (black squares), homogeneous ISM plus an intermediate extinction on
  young stars (blue triangles) and a extreme extinction on young stars (red
  hexagons). In the model with extreme extinction on young stars, all the
  stars younger than $25$ Myr are almost completely extinguished.} \label{fig:norm}
\end{figure*}

\begin{figure*}
\begin{center}
\includegraphics[width=0.33\textwidth]{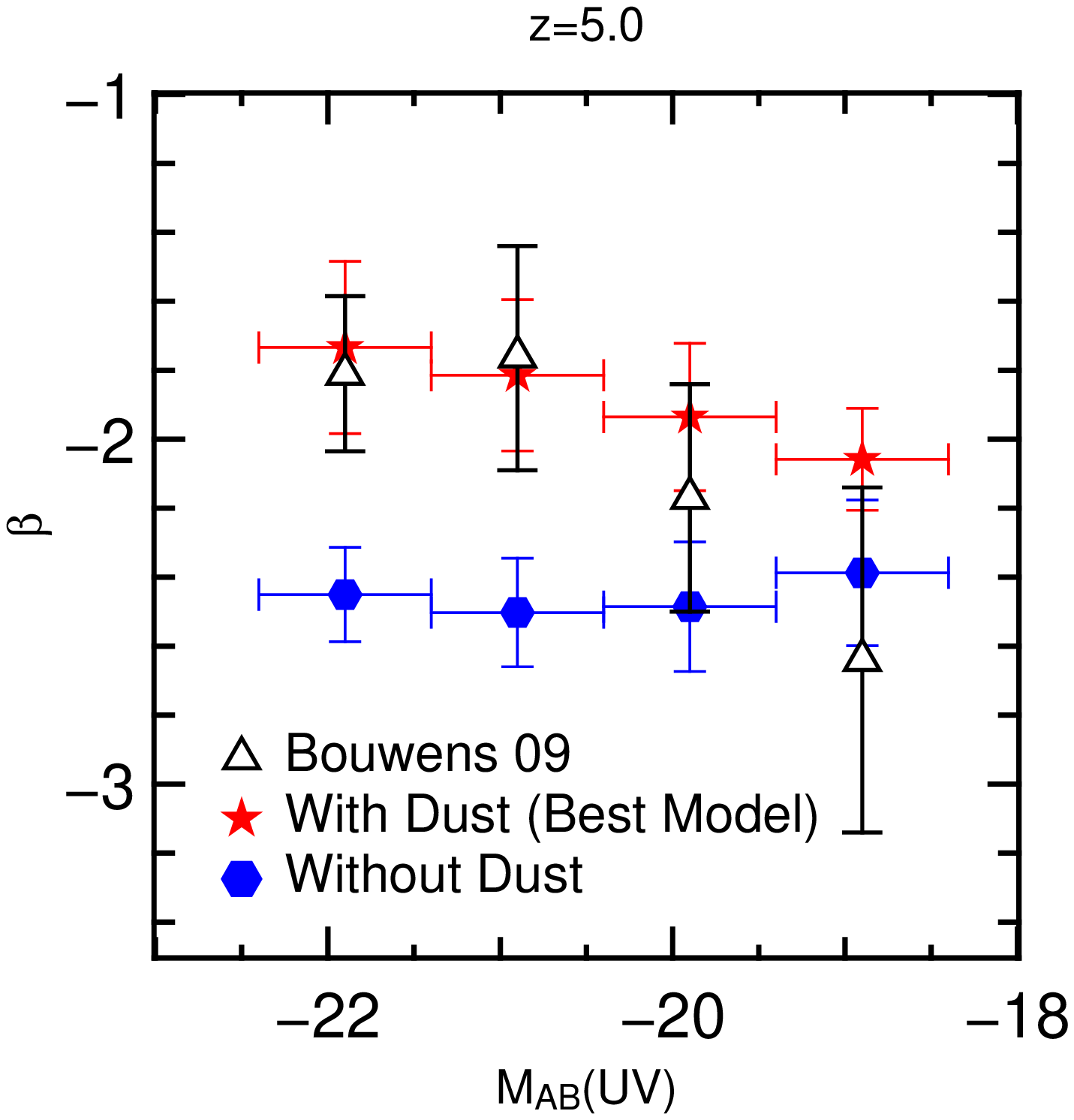}\hspace{-0.9cm}
\includegraphics[width=0.33\textwidth]{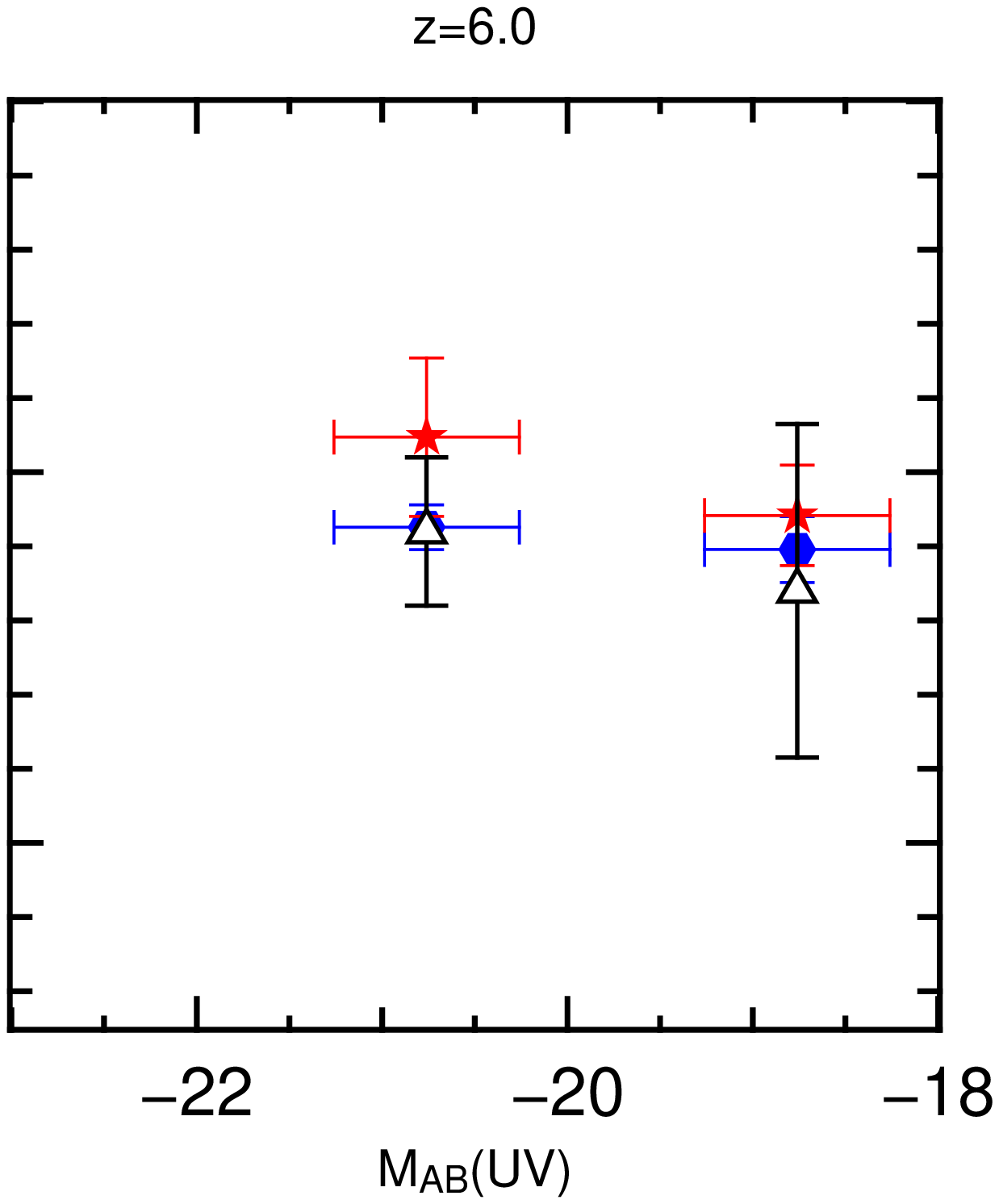}\hspace{-0.9cm}
\includegraphics[width=0.33\textwidth]{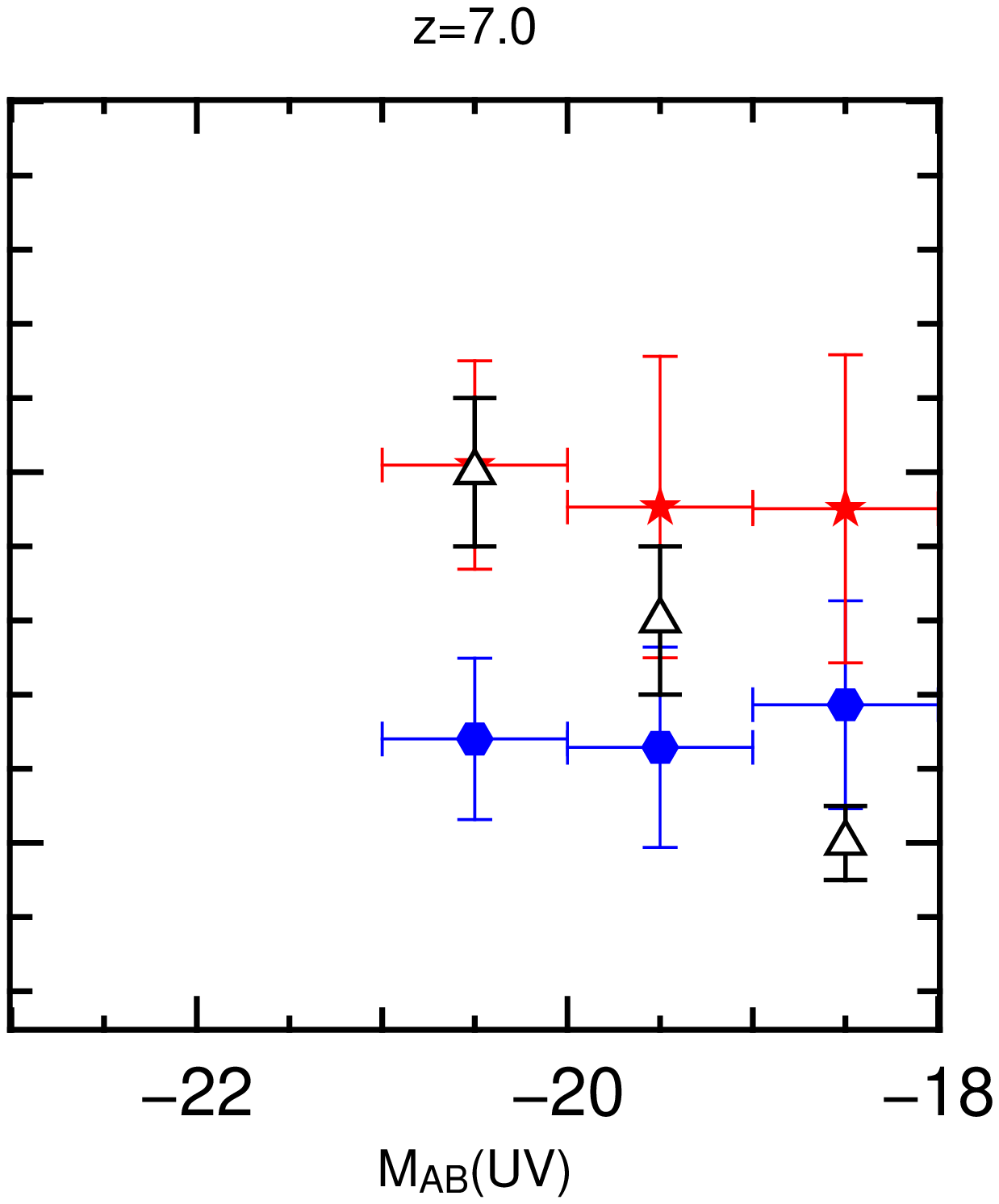}
\end{center}
\caption{The slope, $\beta$,  of the UV-continuum  versus absolute UV
  magnitudes of simulated galaxies  at $z\sim$ $5$, $6$ and $7$.
 The blue points correspond to
  the $\beta$ determinations without extinction, red points correspond
  to the estimation with the extinction model consistent with the UV
  luminosity  functions. The open symbols represent the  estimated values
  of $\beta$ from HST  observations. In both cases the vertical error bars represent a 
  $1$-$\sigma$ scatter.  The brightest galaxies in the simulation
  follows the same  color trend as in  observations. Some discrepancy is still
present at the faint end though. } \label{fig:beta}
\end{figure*}

In the calculation of the luminosity functions we have made first an 
 additional  correction motivated from the different abundance of dark
 matter halos   between the cosmology used in the  simulation (WMAP1,
 $\sigma_8=0.90$ \citep{2003ApJS..148..175S}) and the values of the 
 cosmological parameters estimated from the latest CMB results
 (WMAP5, $\sigma_8=0.796$ \citep{2009ApJS..180..306D}). 
At a given redshift and mass range, the WMAP1 universe has a larger 
  halo number density  than in the WMAP5 one.  
  We estimate the influence of a  different
 cosmology using the Sheth-Tormen formalism \citep{sheth-tormen}.  The
 ratio between the WMAP5 and WMAP1  halo mass functions is used as a
 weight factor for each galaxy (depending on the mass of the host
 halo) in the construction of the LF.  This procedure is justified because
 of the strong correlation shown 
between the UV luminosity and the halo  mass.

 We have  tested  this correction by running  two additional
 simulations with exactly the same mass and force resolutions but  a
 corresponding  smaller volume  to speed up the calculations (i.e  $2\times 256^3$  particles in  a $12.5$ \hMpc comoving boxsize). They were
  performed with same initial conditions using the WMAP1 and WMAP5
 cosmological parameters.   Though not shown here, we
do reproduce the results of the WMAP5 run when using the WMAP1 run with
the aforementioned correction.

The results after correcting for the halo abundance are shown in Figure \ref{fig:norm}. The black squares represent the results without any dust extinction. All the rest-frame UV LFs, without dust extinction,  strongly  over-predict
(up to 2 orders of magnitude) the observational estimates in all the
redshift range, in fairly agreement with results from other $\Lambda$CDM simulations \citep{2006MNRAS.366..705N}.

This discrepancy cannot be attributed to the selection of the Salpeter
IMF to compute the SEDs. Other IMFs, such as Chabrier
\citep{2003PASP..115..763C} or Kroupa \citep{2001MNRAS.322..231K} have
lower mass to luminosity ratios in the UV (at most $\sim 1.5$)  with
respect to a Salpeter IMF,  for a young starburst of a given
age. \citep{2003PASP..115..763C,2003MNRAS.344.1000B}. These IMFs would
then   increase the discrepancy between our simulation and
observations.

One could argue that the excess of galaxies in our simulation
could be related with some of  the  astrophysical processes that shape
the luminosity function at low  redshift   \citep{2003ApJ...599...38B}. 
In particular, an improved modelling of  supernovae and quasar feedbacks
can modify the luminosity function. This scenario cannot be  ruled out
on the basis of the present simulation, and certainly deserves further
study.  Nevertheless, in our case, we want to show that a simple,
physically motivated extinction model can explain the different
behaviour of the simulated and observed UV LF.

If we apply only the extinction due to  the homogeneous ISM, parametrized by $\mu=1.0$ (for any value of $\alpha$) then only the most massive galaxies 
are effectively extinguished.  If we modify the values of $\alpha$
in such a way as to match the bright end of the observed LF, we still find  an excess of simulated galaxies at the faint end. To modify the luminosities at the faint end, specially at $z=5$ and $z=6$, one needs to apply a larger extinction on the young stars.  The correction on the population of young stars turns out to be independent on the total mass of the galaxy, modifying both the bright and faint ends of the LFs.

We explore the parameter space in order to find the best set of
parameters ($\alpha$, $\mu$, $t_c$) in our  extinction model which
provides a good fit to the observational data. In Figure
\ref{fig:norm} the red hexagons show the results for the best fit
parameters: $\alpha=1.5$, $\mu=0.01-0.03$ and $t_c=25$Myr. It means that
stars younger than $t_c=25$Myr are almost completely
extinguished. Thus, the optical depth in the clouds embedding the
young stars must be between $30$ ($\mu=0.03$) and $100$ ($\mu=0.01$)
times larger than the optical depth in the ISM. For comparison,  the
largest values found in local galaxies by \citet{2004MNRAS.349..769K}
are  $\sim 10$ times the optical depth in the ISM.   Furthermore, we
do not find any strong degeneracy between the parameters $\alpha$ and
$\mu$.

We consider now the colors of the galaxies. One of the simplest ways
to observationally estimate the dust attenuation is through the UV slope $\beta$
($f_\lambda\propto \lambda^{\beta}$). In Figure \ref{fig:beta} we show
the values for $\beta$ in the simulated galaxies and  compared to
recent observational estimations \citep{2009arXiv0909.4074B,2009arXiv0910.0001B}.  We find that that the extinguished colors are consistent with the constraints worked out from the observations. There is a minor discrepancy at the faintest
magnitudes (M$_{AB}\sim$ -19) which is not highly significant if we consider
that the error bars in the Figure \ref{fig:beta} represent a $1$-$\sigma$
scatter.  Remarkably, some extinguished simulated galaxies at $z\sim
7$ (not shown in the Figure) still show very blue slopes $\beta\sim 3$
as seen in the observations.

\section{Conclusions}
\label{sec:conclusions}

We have used the {\em MareNostrum High-z Universe}, currently the
largest SPH simulation of galaxy formation, with more than 2 billion
particles, to study the evolution of the rest-frame UV luminosity
function between $5< z <7$. We find that the  UV LF from the
simulation over-predicts systematically the observations. Using a
Chabrier, Kroupa or Top Heavy IMF, instead of the Salpeter IMF we have
used here, will increase the disagreement.

We further correct the luminosity of each galaxy by dust attenuation. If we
correct only by the attenuation produced by an homogeneous ISM, with an
optical depth calculated using Equation \ref{eq:ISM}, we find that the results are
unsatisfactory. In that kind of extinction model, the dust optical depth
turns out to be dependent on the mass of the galaxy, making the bright end of
the LF more extinguished than the faint end, the latter remaining basically
unchanged.

We use then a simple but physically motivated model based on 
\cite{2000ApJ...539..718C}, where the extinction is described by a homogeneous
ISM component plus a clumpy component affecting the younger stars. 
We find that the additional extinction on the younger stars is necessary to
reproduce the overall normalization and shape of the luminosity functions
between $5<z<7$. The final modification on the UV luminosities is independent
on the total mass of the galaxy, meaning that the faint end of the LF can be
modified as well. 

We find that all the stars younger than $25$ Myr must be
obscured almost completely in order to get  a good agreement to the observed
evolution of the LFs. This means that the optical depth in the clouds embedding
the young stars must be between $30$ and $100$ times larger than the optical
depth from the homogeneous ISM. Compared to observations of local galaxies
this proportion between the two optical depths is one order of magnitude
larger than the highest estimates. The agreement of our model to available observational constraints hints for a clumpy ISM at high redshift.

The extinction model we propose is consistent with the observational
constraints of the UV slope between $5<z<7$. We even find some
galaxies with  very blue UV slopes as seen in galaxies at $z\sim 7$
\citep{2009arXiv0910.0001B}. The agreement is less satisfactory at
lower UV magnitudes $M_{AB}\sim -19$. The extinction model we apply to
the very young stars is thus not visible as a strong change in the UV
slopes, but could be observationally detected in the rest-frame
infrared with instruments like HERSCHEL and ALMA.  A complementary
view of the results presented so far will be provided in an upcoming
publication on the physical nature of Lyman-$\alpha$ Emitters in the
{\em MareNostrum High-z Universe}.

\section*{Acknowledgments}
The simulation used in this work is part of the MareNostrum Numerical
Cosmology Project at the BSC. The data analysis has been performed
at the NIC J\"ulich and at the LRZ Munich. We acknowledge the LEA Astro-PF 
collaboration and the ESF  ASTROSIM  network  for  financial support. 
J.E.F-R., S.G. and A.K. acknowledge support by DAAD through the PROCOPE
program. A.K. acknowledges support from ANR-06-BLAN-0172. G.Y. acknowledges support from MEC  projects FPA2006-01105  and AYA2006-15492-C03. 

\bibliographystyle{mn2e}

\end{document}